\documentstyle[prd,aps,floats,tabularx,epsfig]{revtex}
\newcommand{\be}{\begin{equation}}
\newcommand{\ee}{\end{equation}}
\newcommand{\bea}{\begin{eqnarray}}
\newcommand{\eea}{\end{eqnarray}}

\begin{document}
\setlength{\unitlength}{1mm}
\twocolumn[\hsize\textwidth\columnwidth\hsize\csname@twocolumnfalse\endcsname
\title{The Inflationary Gravity Waves in light of recent
Cosmic Microwave Background Anisotropies data.}
\author{Alessandro Melchiorri$^\flat$ and Carolina J. \"Odman$^\sharp$}
\address{ 
$^\flat$ Astrophysics, Denys Wilkinson Building, University of Oxford, 
Keble road, OX1 3RH, Oxford, UK\\
$^\sharp$ Astrophysics Group, Cavendish Laboratory, Cambridge University, Cambridge, U.K.\\}
\maketitle

\begin{abstract}

One of the major predictions of inflation is the existence of a 
stochastic background of cosmological gravitational waves (GW).
These gravitational waves can induce significant temperature
anisotropies in the Cosmic Microwave Background (CMB) 
on the angular scales recently probed by the Archeops 
experiment.
Here, we perform a combined analysis of 
Archeops together with information from other CMB 
experiments and/or cosmological datasets, in order to constrain
the amplitude of the GW background. We find that, for a scale-invariant
GW background, the ratio
of tensor/scalar perturbations at the CMB quadrupole 
is now constrained to be $r \leq 0.43$ 
at $95 \%$ c.l., while the bound on the spectral index of primordial
density fluctuations is $n_S=0.97_{-0.12}^{+0.10}$. 
We discuss the implications for future GW detections through CMB
polarization measurements.
\end{abstract}

\bigskip]

\section{Introduction}

The last years have seen spectacular advances in 
our ability to confront
the inflationary scenario of structure formation 
to observational data.
The ``multiple peaks'' observed in the Cosmic Microwave
Background (CMB) angular power spectrum (\cite{netterfield}, 
\cite{halverson}, \cite{lee}, \cite{pearson}, 
\cite{scott}) are indeed providing strong supporting evidence 
for the inflationary predictions of a flat universe and 
of a primordial background of scale-invariant adiabatic 
perturbations (see e.g. \cite{wang},\cite{carolina}).
More recently, the new CMB results from the Archeops experiment 
(\cite{benoit}) have confirmed and refined the present 
observational status, sampling angular scales between those 
probed by the COBE satellite and the latest high precision datasets.
Again, flatness, adiabaticity and scale invariance 
are in agreement with the data (\cite{benoit2}). 

It has been argued that the next and probably most conclusive
evidence for inflation would be the detection of a stochastic 
background of Gravity Waves (GW) 
(see e.g. \cite{turner}, \cite{dodelson}).
Two types of spacetime metric fluctuations are indeed 
naturally produced during inflation: density perturbations 
({\it scalar} modes), which form the ``seeds'' of structure formation,
and gravity waves ({\it tensor} modes) (\cite{gw}).

The GW background, if detected, would also provide valuable information
on the inflationary scenario. In particular, in most inflationary 
models (and certainly in the simplest ones), the amplitude of the 
GW background is proportional to the square of the energy scale 
of inflation (see e.g. \cite{krauss}). 
Furthermore, a complementary measurement of the 'tilt'
of the GW perturbations (and of the scalar as well) 
can give direct information up to the second derivatives of 
the inflaton potential, sheding light 
on the physics at $\sim 10^{16} GeV$ (see e.g. \cite{hoffman}).

The GW background leaves an imprint on the CMB anisotropies 
at large scales through the Sachs-Wolfe effect.
On scales smaller than the horizon at recombination, however,
 unlike the anisotropies generated by scalar fluctuations,
those generated by GW damp like fluctuations in a fluid of
massless bosons (see e.g. \cite{crittenden}).
Since the theoretical spectrum, normalized to COBE, 
is a linear sum of the scalar and tensor components, 
if there is a relevant contribution from GW 
this would lower the predicted amplitude of the acoustic peaks on 
sub-degree angular scales.

With the advent of the new CMB peaks detections, 
many authors have therefore addressed the question of 
the GW's contribution 
(see e.g. \cite{melk99}, \cite{kmr},
\cite{wang}, \cite{efstathiougw}, \cite{lewisbridle}, 
\cite{percival}, \cite{leach}).
However, despite the different scale dependence, 
robust constraints on tensor modes remain difficult to obtain. 
The decrease in the amplitude of the 
acoustic oscillations induced by GW can indeed be 
compensated by an increase in one of the unconstrained parameters 
of the model, like, for example, the spectral index of scalar 
fluctuations $n_S$. 
Therefore, some form of 'cosmic degeneracy' arises in the 
tradeoff between these two (and more) parameters 
(see \cite{melk99}, \cite{efstathiougw}) and only weak 
constraints on the GW background were obtained.

In this context, and before more accurate polarization data
become available (see discussion below), 
the new results on intermediate angular scales,
as recently provided by Archeops, can offer an interesting opportunity.

As we illustrate in Fig.1, this spectral region has a particular 
sensitivity to a GW contribution. In the figure, we plot
two theoretical power spectra. The models have identical power 
on sub-degree scales and on COBE scales (considering cosmic variance), 
but different tensor contributions,
parametrized by a tensor over scalar ratio of the angular power spectrum 
quadrupole $r=C_2^T/C_2^S$ (see e.g. \cite{kmr}).

As we can see, while the two models are degenerate on scales 
$\ell \ge 200$, the degeneracy is broken on larger angular scales
(see the bottom panel), mostly in the region sampled by Archeops.
Both increasing $n_S$ and adding tensors change 
the {\it rate} of growth of the scalar modes from the Sachs-Wolfe plateau 
towards the first peak and this can in principle be used to 
constrain the GW background.

It is therefore extremely timely to analyze the Archeops data allowing
the possibility of a GW contribution in order to see if the amplitude 
of this background can now be better constrained than in the past.

Furthermore, the GW background produces a unique 
statistical signature in the polarization of the CMB by inducing 
a {\it curl} component (\cite{szpol}, \cite{kks}), often defined as $B$ 
mode, while scalar (but also tensor) perturbations produces a gradient 
component ($E$ mode).
Given the large number of future and ongoing CMB polarization experiments,
it is interesting to forecast from the present CMB temperature 
data the expected  amplitude of the $B$ modes and/or if the 
$E$ modes produced by tensors
can be distinguished from those produced by scalar perturbations 
only.

We pursue this investigation in the present {\it Rapid Communication}
as follows: in Section II we illustrate our 
analysis method. In section III we present our results.
Finally, in section IV, we discuss our findings.

\begin{figure}
\begin{center}
\includegraphics[scale=0.30]{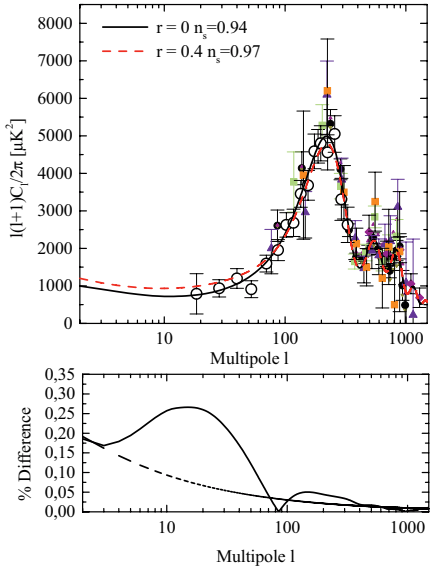}
\end{center}
\caption{Best-fit models to recent CMB data with and without
GW contribution (Top Panel). The Archeops data points are shown
as open circles. In the Bottom panel we plot the $\%$ difference between
the two degenerate models together with the cosmic variance limit
(dashed line) averaged in bins of $\Delta \ell =10$.}
\label{fig1}
\end{figure}

\section{Analysis: Method}

As a first step, we consider a template of flat, adiabatic, 
$\Lambda$-CDM scalar and tensor spectra computed with CMBFAST 
(\cite{sz}), sampling the various parameters as follows:
$\Omega_{cdm}h^2\equiv \omega_{cdm}= 0.05,...0.25$, in steps of  $0.02$;  
$\Omega_{b}h^2\equiv\omega_{b} = 0.009, ...,0.024$, 
in steps of  $0.003$, $\Omega_{\Lambda}=0.5, ..., 0.95$, 
in steps of  $0.05$.
Our choice of the above parameters is motivated by
the Big Bang Nucleosynthesis bounds on $\omega_b$ 
(both from $D$ \cite{burles} and $^4He+^7Li$ \cite{cyburt}),
from supernovae (\cite{super1}) and galaxy clustering
observations (see e.g. \cite{thx}).

Variations in the tensor and scalar spectral indices,
 $n_S$ and $n_T$ are not computationally relevant.
However, we restrict our analysis to relevant inflationary values
$n_S=0.7,...,1.3$ and we fix $n_T=0$ (see discussion below
for different values of $n_T$).

Furthermore, the value of the Hubble constant is not an independent 
parameter, since $h=\sqrt{{(\omega_{cdm}+\omega_b)} / {(1-\Omega_{\Lambda})}}$.
We also include the further top-hat prior $h=0.7\pm0.2$ (\cite{freedman})
and we consider only models with age $t_0>11$ Gyrs.

We allow for a reionization of the intergalactic medium by
varying the compton optical depth parameter 
$\tau_c$ in the range $\tau_c=0.0,...,0.45$ in steps of $0.05$.
We note here that high values of $\tau_c$ 
are in severe disagreement with recent estimates of 
the redshift of reionization $z_{re}\sim 6 \pm 1$ 
(see e.g. \cite{gnedin}) which points towards $\tau_c \sim 0.05-0.10$.
On the other hand, if the reported CBI excess at $\ell \sim 3000$
is due to Sunyaev-Zeldovich effect, then this would favour
values $\tau_c \sim 0.3$ (\cite{bond}).

For the CMB data, we use the recent results from the 
BOOMERanG-98, DASI, MAXIMA-1, CBI, VSA and Archeops experiments. 
The power spectra from these experiments were estimated in  
$19$, $9$, $13$, $14$, $10$ and $16$ bins respectively
(for the CBI, we use the data from the MOSAIC configuration, \cite{cbi}), 
spanning the range $2 \le \ell \le 1500$.
We also use the COBE data from the RADPACK compilation (\cite{radpack}).

For the CBI, DASI, MAXIMA-I and VSA experiments 
we use the publicly available correlation matrices and window functions. 
For the Archeops and BOOMERanG experiments we assign a flat interpolation  
for the spectrum in each bin $\ell(\ell+1)C_{\ell}/2\pi=C_B$,  
and we approximate the signal $C_B$ inside 
the bin to be a Gaussian variable.
The likelihood for a given theoretical model is defined by 
 $-2{\rm ln} {\cal L}=(C_B^{th}-C_B^{ex})M_{BB'}(C_{B'}^{th}-C_{B'}^{ex})$ 
where  $M_{BB'}$ is the Gaussian curvature of the likelihood  
matrix at the peak.

We consider $5 \%$, $10 \%$, $4 \%$, $5 \%$, $3.5 \%$  and $5 \%$ 
Gaussian distributed  
calibration errors (in $\Delta T$) 
for the Archeops, BOOMERanG-98, DASI, MAXIMA-1, VSA, 
and CBI experiments respectively and we include the beam uncertainties 
by the analytical marginalization method presented in (\cite{bridle}).

Finally, we parametrize the GW contribution by the tensor over
scalar quadrupole ratio $r=C_2^T/C_2^S$ and we rescale the
sum spectrum by a prefactor $C_{10}$, assumed to be a free parameter,
in units of $C_{10}^{COBE}$.

\begin{figure}
\begin{center}
\includegraphics[scale=0.40]{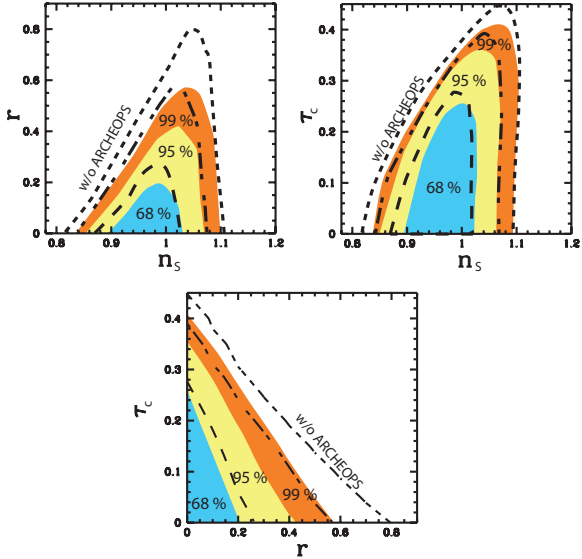}
\end{center}
\caption{$68 \%$, $95 \%$ and $99 \%$ confidence regions in
the $n_S-r$ (Top Panel, Left), $n_S-\tau$ (Top Panel, Right), $r-\tau$
(Bottom Panel) planes for the models considered in our analysis (see text).
The line contours are confidence levels without the Archeops
data.}
\label{fig2}
\end{figure}

\section{Analysis: Results}

The main results of our analysis are plotted in Fig.2.
In the left top panel we plot the likelihood contours in the
$n_S-r$ plane, maximizing over the remaining
{\it nuisance} parameters. As we can see, in the 
framework of models we considered, the gravitational
wave contribution is constrained to be $r \le 0.2$
 ($r \le 0.43$) at $68 \%$ C.L. ($95 \%$ C.L.), with
$n_S=0.97_{-0.07}^{+0.06}$ ($68 \%$ C.L.).
While the inclusion of the Archeops data has little
effect on $n_S$, it drastically improves the constraint
on $r$. Removing the Archeops data yields
$r \le 0.6$ at $95 \%$ C.L..

In the right top panel of Fig.2, we plot the likelihood
contours in the $n_S-\tau_c$ plane.
As we can see, the present CMB constraint on $\tau_c$ is
rather weak, with $\tau_c <0.25$ ($\tau_c<0.36$) at 
$68 \%$ C.L. ($95 \%$ C.L.).
It is interesting to note that the inclusion of the ARCHEOPS
datapoints has little effect.

Finally, in the bottom panel of Fig.2, we plot the
likelihood contours in the $r-\tau_c$ plane.
An increase in $\tau_c$ or $r$ produces a similar damping
on the small/intermediate angular scales.
It is interesting to notice that the present data is allowing
just a well defined amount of small-scale damping. 
Values of $\tau_c \sim 0.3$ are in disagreement with 
the presence of a tensor component.
If $\tau_c >0.2$ then $r <0.05$ at $68 \%$ C.L..

\begin{figure}
\begin{center}
\includegraphics[scale=0.30]{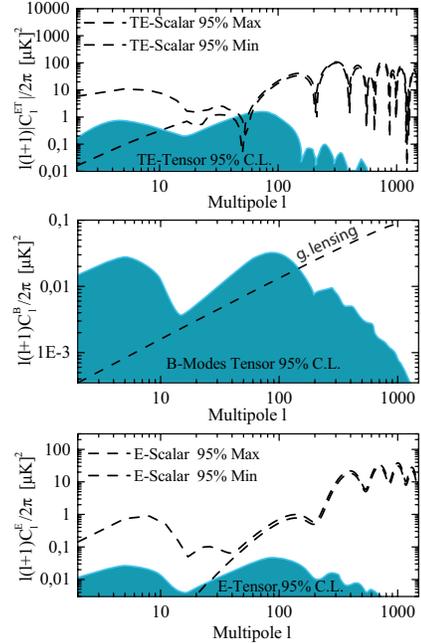}
\end{center}
\caption{Maximum and minimum levels of
temperature-polarization cross correlation (Top Panel),
$B$-modes (Central Panel), $E$-modes (Bottom Panel)
allowed at $95 \%$ C.L. from present CMB temperature data
under the assumption of the models described in the text.}
\label{fig3}
\end{figure}

To each theoretical model in the likelihood planes
produced in Fig.2, is possible to associate a
theoretical polarization power spectrum and translate the 
confidence contours into an expected maximum and minima 
polarization signal.

We do this in the $3$ panels of Fig.3, where we plot
the envelope of the minima and maxima polarization spectra
that, in the panels of Fig.2, are at $95 \%$ c.l.
consistent with the CMB temperature data.

As we can see from the center panel 
of Figure $3$, the level of the $B$-modes, is expected to be 
of $\sim 0.2$ $\mu K$, at maximum. The signal 
is out of the reach of most of the 
current polarization experiment like DASI or POLATRON 
which are sensitive to few $\mu K$.
Near future experiments like B2K or QUEST, will probably have 
enough sensitivity to have a statistical B-mode detection.
However, the $B$-signal in the angular region sampled by these
experiments ($\ell >50$), can be contaminated by a foreground
component due to the conversion of $E$ modes to $B$ modes from
gravitational lensing (see Fig. 3) (\cite{zaldlens}).
Higher-order correlations will be necessary to 
map the cosmic shear and subtract this contribution to the $B$ mode
(\cite{kesden}).

Tensor perturbations produce $E$ modes as well.
However, the amplitude of the $E$ tensor modes is predicted to be
generally much smaller than those from the scalar modes
(see bottom panel). A window of opportunity may appear in the
temperature-polarization
($<TE>$) cross-correlation spectra, where, at $\ell \sim 50$, 
the amplitude from
tensor can be larger than those from scalar modes, leaving a
possible detectable excess for experiments like QUEST or B2K.

In order to cross-check if any information can be obtained
on $n_T$ we performed the analysis on just one cosmological model defined by
$\Omega_{\Lambda} = 0.7$, $\omega_b = 0.022$, $\Omega_{tot} = 1$,
$\tau_c = 0.04$.
We then considered tensor contributions by varying the scalar and
tensor spectral indices independently: $n_S=0.7,...,1.3$ and
$n_T=-0.3,...,0.0$, step $0.01$. 
We found that the tensor spectral index is not constrained by
the present data, but that a value of $n_T = 0$ is preferred. 


\section{Conclusions}

In this {\it Rapid Communication} we have presented new constraints
on the stochastic background of gravitational waves from
recent microwave anisotropy data.
Thanks to Archeops, our results improve the constraints
on tensor modes from previous analyses (see e.g. \cite{percival},
\cite{kmr}).

In the framework of models we considered, we found
(at $95 \%$ C.L.) $r < 0.43$ and $n_S=0.97_{-0.12}^{+0.10}$.
The energy scale of inflation $E_{inf}$ can be related
to tensor by $E_{inf}^4=0.65C_2^Tm_{Pl}^4$.
The above bound translates therefore in 
$E_{inf} \le 1.6 \times 10^{16} GeV$.

When comparing with the results presented in
\cite{benoit2}, a part from the different template of theoretical models
considered, our analysis differs mainly in the
following points: we assumed the low-$\ell$ Archeops bins as
gaussian distributed, we included the COBE data using the
RADPACK compilation, we have a strong upper limit
on $\omega_b <0.025$ from BBN and, finally, we numerically computed the
models with $\tau_c>0$ (while in \cite{benoit2} an analytical 
formula was used).

The GW background induces a unique signature in the polarization
of the CMB by producing a curl component, not present in the
case of scalar perturbations.
In the set of models we considered (and under the assumption
of a bayesian method of statistical analysis) we found that
the maximum expected level of $B$ modes allowed by current data is 
of about $\sim 0.2 \mu K$, which can be partially attainable by near future
experiments and severly contaminated by lensing
$E \rightarrow B$ conversion.

The $E$ modes expected from gravity waves are lower than
the $E$ modes expected from scalar perturbations. However, the tensor
$<TE>$ cross-correlation might be larger at $\ell \sim 50$.

In this context, deviations in the $<TE>$ cross-correlation
scalar spectrum at $\ell \sim 50$ can possibly offer competitive
information with respect to $B$ modes search.

All our predictions derived from the temperature data
are consistent with the recent claim of detection 
of polarization $E$ modes from the DASI experiment 
(\cite{newdasi}).

As a final remark, we want to stress that the results presented
here have been obtained under the assumption of a theoretical
framework. The bounds on the polarization spectrum must be considered 
just as an indication of what future observations may detect.

In particular, we just considered $n_T=0$ in the main analysis
and we looked at the effect of having $n_T$ as low as $-0.3$. 
For inflation, only values $n_T \le 0$ 
can be considered and we checked that varying $n_T$
 has little effect on the final results.
 Spectra with ``blue'' ($n_T>0$) spectral indices can be produced
in Ekpyrotic (\cite{steinhardt}) or Pre-Big Bang (see e.g.
\cite{melchiorri}) scenarios. However,
extremely blue spectra ($n_T \sim 2$) are excluded
by constraints on the GW energy density background from
timing milli-second binary pulsars \cite{pulsar}.
Allowing for extra primordial perturbation modes like
isocurvature, will probably tight our constraints on GW, since
the shape of CDM scalar isocurvature modes is similar
to those from adiabatic tensor modes. However, considering
the most general initial conditions scheme and
including cross correlations, will certainly enlarge our constraints
(\cite{bmt}).
Including curvature ($\Omega_{tot} \neq 1$) would relax our 
bounds on $r$ (see e.g. \cite{wang}). Non-flat models in agreement with 
the CMB data are in general closed models, which, a part from a few 
exceptions (\cite{linde}), are difficult to obtain from inflation.
Finally, including a different model for dark energy like
quintessence would change the large scale anisotropy through the
Integrated Sach Wolfe effect (see e.g. \cite{bean}),
affecting our constraints as well.

Even if the results presented here do not hint for a presence
of GW background, the data is still consistent with a sizable
tensor contribution. It will therefore be the duty of future and ongoing
experiments to scrutinize this fundamental prediction from inflation.

{\bf Acknowledgements}

We wish to thank Anthony Challinor, Asantha Cooray,
Will Kinney, Rocky Kolb, Mike Hobson, 
Robert Izzard, Anthony Lasenby, Antonio Riotto and Francois Xavier-Desert.
AM is supported by PPARC. 
CJO is supported by a Girton College Scholarship and an 
Isaac Newton Studentship.

\end{document}